# MEDICAL SCIENCES

## Human cancers over express genes that are specific to a variety of normal human tissues

Joseph Lotem*†, Dvir Netanely†‡, Eytan Domany‡§ and Leo Sachs*§

Departments of Molecular Genetics* and Physics of Complex Systems‡, Weizmann Institute of Science, Rehovot 76100, Israel

Editorial correspondence and proofs should be addressed to:

Professor Leo Sachs, Department of Molecular Genetics, Weizmann Institute of Science, Rehovot 76100, Israel; Tel. 972-8-934-4068; Fax. 972-8-934-4108; E-mail
leo.sachs@weizmann.ac.il

† JL and DN contributed equally to this work

§ To whom correspondence should be addressed: E-mail: eytan.domany@weizmann.ac.il;
leo.sachs@weizmann.ac.il

We have analyzed gene expression data from 3 different kinds of samples: normal human tissues, human cancer cell lines and leukemic cells from lymphoid and myeloid leukemia pediatric patients. We have searched for genes that are over expressed in human cancer and also show specific patterns of tissue-dependent expression in normal tissues. Using the expression data of the normal tissues we identified 4346 genes with a high variability of expression, and clustered these genes according to their relative expression level. Of 91 stable clusters obtained, 24 clusters included genes preferentially expressed either only in hematopoietic tissues or in hematopoietic and 1-2 other tissues; 28 clusters included genes preferentially expressed in various non-hematopoietic tissues such as neuronal, testis, liver, kidney, muscle, lung, pancreas and placenta. Analysis of the expression levels of these 2 groups of genes in the human cancer cell lines and leukemias, identified genes that were highly expressed in cancer cells but not in their normal counterparts, and were thus over expressed in the cancers. The different cancer cell lines and leukemias varied in the number and identity of these over expressed genes. The results indicate that many genes that are over expressed in human cancer cells are specific to a variety of normal tissues, including normal tissues other than those from which the cancer originated. It is suggested that this general property of cancer cells plays a major role in determining the behavior of the cancers, including their metastatic potential.

Genetic and epigenetic changes contribute to the development of cancer and lead to abnormalities in regulation of cell viability, multiplication and differentiation (reviewed in refs. 1-3). Such changes in DNA and chromatin structure can lead to aberrant gene expression, and analysis of global gene expression detected changes in gene regulation in different types of cancer (4-9). But are the up-regulated genes in cancer cells limited only to those genes that are normally preferentially expressed in the same tissues from which the cancer originated? We have previously shown that cells of a mouse myeloid leukemia cell line highly express various genes that are normally preferentially expressed in different non-hematopoietic tissues including neuronal, liver, testis and muscle (10). We have now determined whether this phenomenon is common to different types of human cancers, using DNA microarray expression data from normal human tissues (11), different human cancer cell lines (11) and different subtypes of human leukemia from patients (6, 12). Cluster analysis of gene expression in normal human tissues revealed that many genes that are over expressed in different human cancer cells are normally preferentially expressed in tissues other than those from which the cancer originated. The results also indicate that the number and identity of these highly expressed genes were different in different types of cancers.

**Materials and Methods**

**Data sets.** Three DNA microarray data sets were used: In the first data set, mRNA expression levels of genes in normal human tissues and in various human cancer cell lines (11) were measured using 2 DNA microarrays, the Affymetrix HG-U133A array and the GNF1H, a custom designed array (11). The data set downloaded from Su et al (11) contained 33689 probe sets (PS). We removed all PS that were mapped to more than one gene symbol, leaving 33440 PS that were used for further analysis. The downloaded data set included 72 normal human tissue samples in duplicates and 7 human cancer cell lines also in duplicates (11). The cancer cell lines (11) included the T cell lymphoma MOLT4, the B-cell lymphoma 721, the Burkitt's lymphomas Raji and Daudi, the myeloid leukemia HL-60, the chronic myeloid leukemia derived cell line K562 and the colorectal carcinoma SW480. The two other data sets used included mRNA expression data from leukemic blast cells of 132 pediatric patients with different acute lymphoid leukemia (ALL) subtypes (12), 5 pediatric patients with T-ALL with a rearranged *MLL* gene (6) and 130 pediatric patients with different acute myeloid leukemia (AML) subtypes (6). The ALL subtypes included T-ALL without or with rearrangement of the *MLL* gene and 6 different B-ALL subtypes, including those with a rearranged *MLL* gene, with chromosomal translocations involving BCR/ABL, E2A/PBX1 or TEL/AML1, with a hyperdiploid number of chromosomes (HD50) and others (12). There were 6 different AML subtypes including those with a rearranged *MLL* gene, with chromosomal translocations involving PML/RARα, AML1/ETO or CBFβ/MYH11, M7 megakaryocytic leukemia and others (6). Gene expression in these data sets was measured with the Affymetrix HG-U133A array. For all data sets, the expression value for each gene was determined using the MicroArray Suite version 5.0 (MAS 5.0) software (13).

**Clustering of Highly Variable Genes in Normal Human Tissues.** Expression values of PS in the duplicates of each normal tissue sample were averaged, expression values <20 were adjusted to 20 to eliminate noise from the data and all values were then $\log_{10}$ transformed. The 33440 PS were filtered to select those genes that show a highly variable expression level in the 72 human tissue samples. We used two criteria to filter the PS and those PS that satisfied either criterion, were included: I. High ($\geq 0.4$) standard deviation of the log-transformed expression (LTE), measured over the different human tissues; II. LTE range of at least 2 *and* LTE value of ≥3 standard deviations below or above the mean in at least one tissue. The 4346 highly variable PS that passed either of these two criteria, were clustered according to their expression in the different human tissues, using the coupled two-way clustering (CTWC) algorithm (14). Stable clusters were identified by CTWC after applying the Mean Field Approximation version (15) of the Super Paramagnetic Clustering (SPC) algorithm (16). CTWC was applied using the default parameters, except for a minimal cluster size of 10. Prior to the first clustering step, the LTE of each gene was centered and normalized over the samples used.

**Identification of Highly Expressed Genes.** We used all the available PS to calculate separately the 85[th] percentile of the un-normalized expression values for each of the normal tissue samples and cancer cell lines. In the case of leukemic cells from patients, we first averaged the expression values over all patients with the same leukemia subtype, and took the 85[th] percentile of these average values. All the PS that were expressed at values higher than this threshold were defined as *highly expressed*. We have shown previously that changing the threshold used to define highly expressed genes to the 80[th] or 90[th] percentiles, did not affect the conclusions drawn from the analysis (10).

**Results**

**Clustering of Highly Variable Genes in Normal Human Tissues.** After filtration, the 4346 PS that showed highly variable expression in the 72 different normal tissue samples were clustered by CTWC. This clustering operation yielded 91 stable gene clusters that show *preferential expression* in different tissues (see Fig. 4, which is published as supporting information on the PNAS web site). The term "preferential expression" refers to high relative expression levels of the majority of a cluster's genes in a particular subset of the samples, which was determined by inspection of color-coded expression matrices such as the left panels of Figs. 1 and 2. The genes included in 14 of these clusters showed preferential expression only in hematopoietic tissues. Further sub-classification within the hematopoietic tissues indicated that some of these clusters contained genes preferentially expressed either in T cells (see left panel, Fig 1A), B cells (Fig. 1B), myelomonocytic cells (Fig 1C) or erythroid and bone marrow endothelial cells (Fig. 1D), whereas other clusters contained genes expressed at similar levels in most hematopoietic cell types (Fig. 1E). In addition, 10 other clusters contained genes that were preferentially expressed in hematopoietic tissues plus 1-2 other tissues. For further analysis we shall refer to these 24 clusters (10 hematopoietic tissues only and 14 hematopoietic plus 1-2 other tissues) as hematopoietic (H) clusters (see the list of H clusters in Table 3, which is published as supporting information on the PNAS web site).

In addition to the H clusters, there were 28 clusters that contained genes preferentially expressed in various non-hematopoietic (NH) tissues. Some of these NH clusters, contained genes preferentially expressed in only one type of tissue such as neuronal, testis, placenta (Fig. 2 A-C, respectively, left panels), kidney, adrenal, pancreas or thyroid. Other NH clusters contained genes showing preferential expression in 2-3 tissues such as neuronal and testis, or in ≥4 different non-hematopoietic tissues (Fig. 2 D, left panel and Table 3).

**Testing for Distortion due to Normalization.** The clustering operation that identified the H and NH clusters was based on LTE values that were centered and normalized, for each PS. This step may distort the relative expression levels of the genes in a particular sample. To show that this is not the case, we checked the overlap of the H and NH cluster genes with those that are identified as highly expressed genes, applying our standard 85th percentile threshold on the raw LTE values. The results indicate that 92.5% of the H cluster genes, i.e. 1046 out of 1130, were highly expressed in some hematopoietic tissues (Table 1). In contrast, only 3.5% of the H-cluster genes, 40 genes, were highly expressed in all normal tissues. There was also a low frequency of H-cluster genes that were highly expressed in various non-hematopoietic tissues, and for example, there were 126 such highly expressed genes in appendix (Table 1). An illustration of this phenomenon in two H clusters is shown in the Fig. 3 A and B. Similar to the H clusters, 95.3% of the NH cluster genes, 1533 out of 1609, were highly expressed in the corresponding non-hematopoietic tissues, but only 1.8% of the NH cluster genes, 29 genes, were highly expressed in all normal tissues and only 273 of the NH cluster genes, were highly expressed in hematopoietic tissues (Table 1). An illustration of this phenomenon in two NH clusters is shown in Fig. 3 C and D. These results indicate that the genes included in the various H or NH clusters according to their relative expression profile in different tissues, were also highly expressed mainly in the corresponding hematopoietic and non-hematopoietic tissues.

**Search for Genes that are Highly Expressed in Leukemic Cell Lines and in Some Normal Tissues, but *not* in Normal Hematopoietic Tissues.** To compare gene expression in the normal tissues and cancer cell lines, we standardized the expression values of each gene over all the normal tissues and the 7 cancer cell lines included in the data set (11) (Figs. 1 and 2, left and middle panels, respectively). In these Figures, the internal ordering of the genes within each H and NH cluster was based on hierarchical clustering applied over the cancer cell lines. Thus, within each H or NH cluster, genes with similar expression profile over the cancer cell lines are adjacently placed (for ease of inspection). On the right panels of Figs. 1 and 2 we mark the genes that were highly expressed in the cancer cell lines.

The results with H-cluster genes indicate that different leukemia/lymphoma cell lines varied in the number of highly expressed genes (Table 1). As expected from their hematopoietic origin, almost all the genes that were highly expressed in the different cell lines were also highly expressed in their normal hematopoietic counterparts (Table 1). Those few H-cluster genes that were highly expressed in the leukemia/lymphoma cell lines but not in any of the normal hematopoietic counterparts are listed in Table 4, which is published as supporting information on the PNAS web site. The results also show that only the T cell leukemia Molt4 highly expressed some genes that are preferentially expressed in normal thymus (Fig. 1 A). Furthermore, only B cell lymphomas 721, Raji and Daudi and the myeloid leukemia HL-60 highly expressed some genes that are preferentially expressed in normal B cells (Figs. 1B). In addition, the K562 cell line that can be induced to differentiate along the erythroid lineage (17), showed the highest number of highly expressed genes that are preferentially expressed in normal erythroid precursor cells (Fig. 1D). These results indicate that the human leukemia cell lines maintain certain features that characterize their normal cell lineage.

The behavior described above is the standard and expected one, high expression of genes in cancer cell lines is accompanied by high expression in the normal tissue of origin. We turned to search for a more interesting, non-standard expression pattern, that of genes that are highly expressed in some cancer cell lines, low in the normal tissue from which these cancers originated, but high in some other normal tissue. Here we demonstrate that this non-standard scenario is indeed observed, concentrating first on leukemia/lymphoma cell lines. First, we noted that the NH-cluster genes exhibit variability in their expression over the different leukemia/lymphoma cell lines. For each of these different cell lines we identified the highly expressed genes that belong to the NH clusters. The leukemia/lymphoma cell lines highly expressed some NH cluster genes that are preferentially expressed in normal neuronal, testis, placenta (Fig. 2), liver, kidney, thyroid, lung and some others. The different leukemia/lymphoma cell lines varied in the number of the highly expressed NH cluster genes (Table 1). For each cell line we calculated the fraction of its highly expressed NH genes that were not highly expressed in any of the normal hematopoietic tissue samples. This fraction represents genes that are over expressed in the cell lines and it varied from 20% to 48% in different cell lines (Table 1 and see the list of genes in Table 5, which is published as supporting information on the PNAS web site 5). Note that of the NH-cluster genes that were over-expressed in the cell lines, almost 80% were over expressed only in a single leukemic cell line (Table 5). These results indicate that different human leukemia/lymphoma cell lines over-express genes that are normally preferentially expressed in tissues other than the hematopoietic tissue from which the leukemias originated.

**Identification of Genes that are Over Expressed in Leukemic Cells from Human Patients with Different Subtypes of Lymphoid or Myeloid Leukemia.** Cancer cells from patients with various types of cancer including leukemia over express various genes compared to their normal tissue of origin (4-9). In view of our previous results (10) and the results described above with leukemia/lymphoma cell lines, we now determined the extent to which leukemic cells from patients with different leukemia subtypes also highly express genes that are normally preferentially expressed in non-hematopoietic tissues. The gene expression data used in normal human tissues (11), leukemic cells from pediatric patients with ALL (12) or AML (6) were from different data sets, which prohibited direct comparison of gene expression values in these

different studies. Therefore, we first calculated the average expression values of every gene in the data sets from all patients with the same leukemia subtype, and identified the highly expressed genes. We then determined which of the H or NH-cluster genes were highly expressed in the different leukemia subtypes. Of the 1130 PS included in the normal H clusters, 1038 PS were also present in the human ALL and AML data sets. Of these 1038 H-cluster PS, the number of those that were highly expressed in leukemic cells from the different leukemia subtypes varied (over the subtypes) between 244 and 329, and almost all of these PS were also highly expressed in normal hematopoietic tissues (Table 2). In addition, of the 137 genes included in H-clusters G19, G28, G32 and G72 preferentially expressed in normal myelomonocytic cells (Table 3), the T-ALLs highly expressed only 3-5 genes, the B-ALLs 4-12 genes, the M7 megakaryocytic leukemia 12 genes, whereas the other AMLs highly expressed 17-42 genes. Of the 37 H-cluster genes preferentially expressed in normal B cells, the T-ALLs and the different AMLs highly expressed only 1-3 genes whereas the B-ALLs highly expressed 15-17 genes. Thus, like the leukemia/lymphoma cell lines, different human leukemias also showed a high degree of cell lineage fidelity.

Turning again to search for genes with non-standard expression patterns, we analyzed the NH-cluster genes, and found that the different ALL and AML subtypes highly expressed 30-55 genes, which were preferentially expressed in various normal non-hematopoietic tissues (Table 2). Of all the NH-cluster genes that were highly expressed in the different leukemias, 42 genes were not highly expressed in any normal hematopoietic tissue and are thus over expressed in the leukemias (Table 2 and see the list of these genes in Table 6, which is published as supporting information on the PNAS web site). Of these 42 over expressed NH-cluster genes, 30 genes were over expressed only in a single leukemia subtype, and only 4 genes in $\geq 4$ leukemia subtypes (Table 6). These results indicate that like the leukemia/lymphoma cell lines, leukemic cells from patients also over express various genes that are normally preferentially expressed in various non-hematopoietic tissues including neuronal, testis, liver and placenta, and most of these genes were over expressed in just a single leukemia subtype. However, the number of NH cluster genes that were over expressed in the ALLs was higher than in the AMLs, 35 versus 14 genes, respectively, including 13 versus 4 genes preferentially expressed in neuronal and testis tissues (Table 6).

**Identification of Genes that are Over Expressed in SW480 Adenocarcinoma Cell Line.** The ability of leukemic cells to over express genes that are normally preferentially expressed in various non-hematopoietic tissues raised the question whether other types of cancer cells also possess this property. The results indicate that SW480 cells highly expressed various H-cluster genes (Table 1 and Fig. 1 C, D and E), although they had a lower number of such genes compared to the leukemia/lymphoma cell lines (Table 1). Furthermore, 69% of the highly expressed H-cluster genes in SW480 cells, were not highly expressed in the appendix (Table 1), which we used as their normal counterpart, and are thus over expressed in SW480. The H cluster genes that are over expressed in SW480 (Table 4) include the apoptosis inhibitors *SERPINA1* and *BIRC5* and some genes involved in human cancer-associated translocations such as *LMO2*, *RUNX1* and *TCF3* that could play a role in their cancer phenotype. Analysis of genes that are included in various NH clusters showed that SW480 highly expressed more such genes than the leukemic cell lines (Table 1). However, unlike the leukemia/lymphoma cell lines, SW480 did not highly express any of the genes that are preferentially expressed in normal testis (Fig. 2B and Table 5). Of the 162 NH cluster genes that were highly expressed in SW480, 68% were not highly expressed in normal appendix (Table 1). In addition, only 21 of these over expressed NH cluster genes were common to SW480 and at least one of the leukemia or lymphoma cell lines (Table 5). The results indicate that human adencocarcinoma cells, like leukemia/lymphoma cells, over-express many genes that are normally preferentially expressed in tissues other than their tissue of origin. As with the H cluster genes, the list of the over expressed NH cluster genes in SW480 includes many genes that are known to be over expressed in various types of human cancer and could contribute to cancer development and progression including *HOXA9, HOXB6, SOX9, CCND1, EGFR, SERPINE1, KRT8, KRT18, KRT19, TIAM1, FHL2* and *L1CAM* (Table 5).

**Discussion**

Normal hematopoietic stem cells express genes that are preferentially expressed in various normal non-hematopoietic tissues (18). We have previously shown that cells of a mouse myeloid leukemia cell line also highly express genes that are normally preferentially expressed in non-hematopoietic tissues such as neuronal, testis, liver and muscle tissues (10). It is well established that human cancer cells from patients over express various genes compared to their normal tissue of origin (4-9). We have now determined to what extent do different types of human cancer cells over express genes that are normally preferentially expressed in hematopoietic and non-hematopoietic tissues. We clustered genes that showed a highly variable expression level in 72 different normal human tissue samples and selected 2 major cluster categories; H clusters with genes preferentially expressed only in hematopoietic tissues or in hematopoietic tissues plus 1-2 other tissues, and NH clusters with genes preferentially expressed in a single or multiple non-hematopoietic tissues. More than 92% of the genes included in all H or NH clusters were highly expressed in the corresponding normal tissues.

We determined which of the H or NH cluster genes were highly expressed in each of the human cancer cell lines tested and in different human ALL and AML subtypes. The results with H cluster genes indicated that different leukemia/lymphoma cell lines and leukemic cells from ALL and AML patients showed good lineage fidelity. Both the myeloid leukemia HL-60 and the B lymphoma cell lines showed the same number of highly expressed H cluster genes that are preferentially expressed in normal B cells. It is possible that this phenomenon may reflect a common normal hematopoietic precursor cell for both B cells and myeloid cells (19). As expected from the large fraction of H cluster genes that are highly expressed in normal hematopoietic tissues, almost all the H cluster genes that were highly expressed in the leukemia/lymphoma cell lines and leukemia patients were also highly expressed in normal hematopoietic cells. The colon adenocarcinoma SW480 also highly expressed 154 H cluster genes, but 69% of these genes were not highly expressed in normal appendix, which we used as a normal counterpart of SW480, and are thus over expressed in SW480 compared to normal appendix. Two of these H cluster genes that are over expressed in SW480 cells, *SERPINA1/ 1 ANTITRYPSIN* and *BIRC5/SURVIVIN* are anti-apoptotic genes (20, 21). Furthermore, *SURVIVIN* is over expressed in many human cancers (21) including colorectal cancer in which it is regulated by the TCF/βcatenin pathway (22) and contributes to the radiation resistance in SW480 cells (23). Some of the other H cluster genes that are over expressed in SW480 such as *LMO2*, *RUNX1/AML1* and *TCF3/E2A* are involved in human cancer-associated translocations (24-26). Other H cluster genes that are over expressed in SW480, are also over expressed in a variety of human cancers and could contribute to development and progression of cancer due to their functions in regulating cell viability, proliferation, DNA repair, adhesion and invasiveness (Table 7 which is published as supporting information on the PNAS web site).

The results with NH cluster genes indicate that the leukemic cell lines, the ALL and AML leukemia subtypes and SW480 highly expressed various genes that are preferentially expressed in tissues other than those from which the cancers originated including neuronal, liver, kidney, thyroid, lung or placenta. However, unlike the leukemia/lymphoma cells, SW480 did not highly express any of the genes that are preferentially expressed in normal testis. It will be interesting to determine whether other colorectal cancer cell lines share this property. The results have also indicated that a large proportion of the NH cluster genes that are highly expressed in the different cancer cells were over expressed in the cancer cells compared to their tissue of origin. Most of these genes were over expressed only in a single cancer cell line or leukemia subtype, indicating that the different cancer cells show differences both in the number and the identity of their over expressed genes. Many of these genes are up regulated in various types of human cancer and could contribute to cancer development and progression (Tables 5 and 8 which are published as supporting information on the PNAS web site). In addition, it was reported by others that some of these genes including *SOX9, CCND1, EGFR, SERPINE1, TIAM1, FHL2* and *L1CAM* are indeed highly expressed in SW480 cells (27-33). Furthermore, *CCND1* and *L1CAM* are targets of the TCF/β-catenin pathway (28, 33), which is aberrantly activated in various cancers including colorectal cancer from which SW480 cells were derived.

It is suggested that the ability to over express genes that are normally preferentially expressed in tissues other than the cancer's origin, is a general property of cancer cells that plays a major role in determining the behavior of the cancers, including their metastatic potential. The results from the pediatric ALL and AML patients indicate that the ALLs over expressed more NH cluster genes than the AMLs, including genes preferentially expressed in neuronal tissues and testis. It will be interesting to find out whether this phenomenon is associated with the higher frequency of leukemia involvement in the central nervous system in ALL versus AML pediatric patients (34, 35).

In the present study we scored genes that are over expressed in cancer cells using a very stringent threshold requirement, namely, only those genes that are above the 85th percentile in the cancer cells but below this threshold in their normal counterparts. Therefore, all the over expressed genes we scored in cancer cells are also highly expressed genes. It is expected that there are other over expressed genes in cancer cells, whose level of expression in both normal and cancer cells is either above or below the 85th percentile. Our results also indicate that there were differences in the identity of most of the over expressed genes between different cancer cell lines, even between leukemic cell lines from the same lineage. Therefore, the fact that we used the average expression values of genes from all patients with a given leukemia subtype, presumably resulted in detection of only a fraction of over expressed genes, those that are commonly over expressed in many of the patients. It is expected that additional over expressed genes can be identified, which show patient to patient differences even with the same leukemia subtype.

L. S. is the Otto Meyerhoff Professor of Molecular Biology. His work was supported by the Benoziyo Institute of Molecular Medicine, the Dolfi and Lola Ebner Center for Biomedical Research, Dr. and Mrs. Leslie Bernstein, Mrs. Bernice Gershenson. E. D. is the incumbent of the H J Leir Professorial Chair. His work was partially supported by the Ridgefield Foundation and the Minerva Foundation.

**Legends to Figures**

**Fig. 1.** Examples of different hematopoietic (H) clusters. Genes that show highly variable expression level in 72 normal tissue samples were clustered by CTWC as described in *Materials and Methods*. Some clusters showing preferential expression in normal hematopoietic tissues and the relative gene expression levels in 7 human cancer cell lines are shown in the left and middle panels, respectively, according to the color code shown on the left. Highly expressed genes in the cancer cell lines (>85%) are shown in the right panel, marked as red boxes.

**Fig. 2.** Examples of different non-hematopoietic (NH) clusters. Some clusters showing preferential expression in various normal non-hematopoietic tissues and the relative gene expression levels in 7 human cancer cell lines are shown in the left and middle panels, respectively, according to the color code shown on the left. Highly expressed genes in the cancer cell lines (>85%) are shown in the right panel, marked as red boxes. The order of the normal tissue samples and cancer cell lines is as in Fig. 1.

**Fig. 3.** Highly expressed genes in some hematopoietic (H) and non-hematopoietic (NH) clusters in normal tissues. *(A and B)* H clusters; *(C and D)* NH clusters. Genes that are highly expressed in normal hematopoietic tissue samples or in any of the other normal tissues are marked as pale blue or red boxes, respectively. The order of the normal tissue samples is as in Fig. 1.

**Table 1. Number of highly expressed PS in normal tissues and different human cancer cell lines**

### Hematopoietic (H) clusters

| Normal tissues | | | Cancer cell lines | | | | | | |
|---|---|---|---|---|---|---|---|---|---|
| n* | H | Appendix | Molt4 | 721 | Raji | Daudi | HL-60 | K562 | SW480 |
| 1130 | 1046 | 126 | 226 (0) | 378 (7) | 260 (4) | 281 (3) | 247 (2) | 165 (3) | 154 (107) |

### Non hematopoietic (NH) clusters

| Normal tissues | | | | Cancer cell lines | | | | | | |
|---|---|---|---|---|---|---|---|---|---|---|
| n* | H | NH | Appendix | Molt4 | 721 | Raji | Daudi | HL-60 | K562 | SW480 |
| 1609 | 273 | 1533 | 295 | 54 (20) | 70 (34) | 70 (24) | 53 (11) | 46 (17) | 110 (50) | 162 (110) |

* n, total number of PS in all H or NH clusters. Values in brackets are the number of PS that were highly expressed in cancer cell lines but not in their normal counterparts.

**Table 2. Number of highly expressed PS in different subtypes of human ALL and AML**

**Hematopoietic (H) clusters**

| n* | T-ALL | | B-ALL | | | | | |
|---|---|---|---|---|---|---|---|---|
| | +MLL | - MLL | +MLL | BCR/ABL | E2A/PBX1 | TEL/AML1 | HD50 | Others |
| 1038 | 286 (0) | 286 (1) | 265 (0) | 302 (0) | 249 (1) | 300 (0) | 278 (0) | 294 (0) |

| | AML | | | | | |
|---|---|---|---|---|---|---|
| | +MLL | PML/RARα | AML1/ETO | CBFβ/MYH11 | M7 | Others |
| | 318 (1) | 244 (0) | 277 (0) | 329 (0) | 281 (0) | 306 (0) |

**Non hematopoietic (NH) clusters**

| n* | T-ALL | | B-ALL | | | | | |
|---|---|---|---|---|---|---|---|---|
| | +MLL | -MLL | +MLL | BCR/ABL | E2A/PBX1 | TEL/AML1 | HD50 | Others |
| 1450 | 51 (10) | 33 (3) | 50 (9) | 51 (5) | 53 (9) | 49 (13) | 45 (6) | 44 (4) |

| | AML | | | | | |
|---|---|---|---|---|---|---|
| | +MLL | PML/RARα | AML1/ETO | CBFβ/MYH11 | M7 | Others |
| | 44 (4) | 53 (6) | 41 (1) | 35 (2) | 55 (5) | 30 (1) |

*n, total number of PS in all H or NH clusters. Values in brackets are the number of PS that are highly expressed in cells from different leukemia subtypes but not in normal hematopoietic cells.

Fig 1

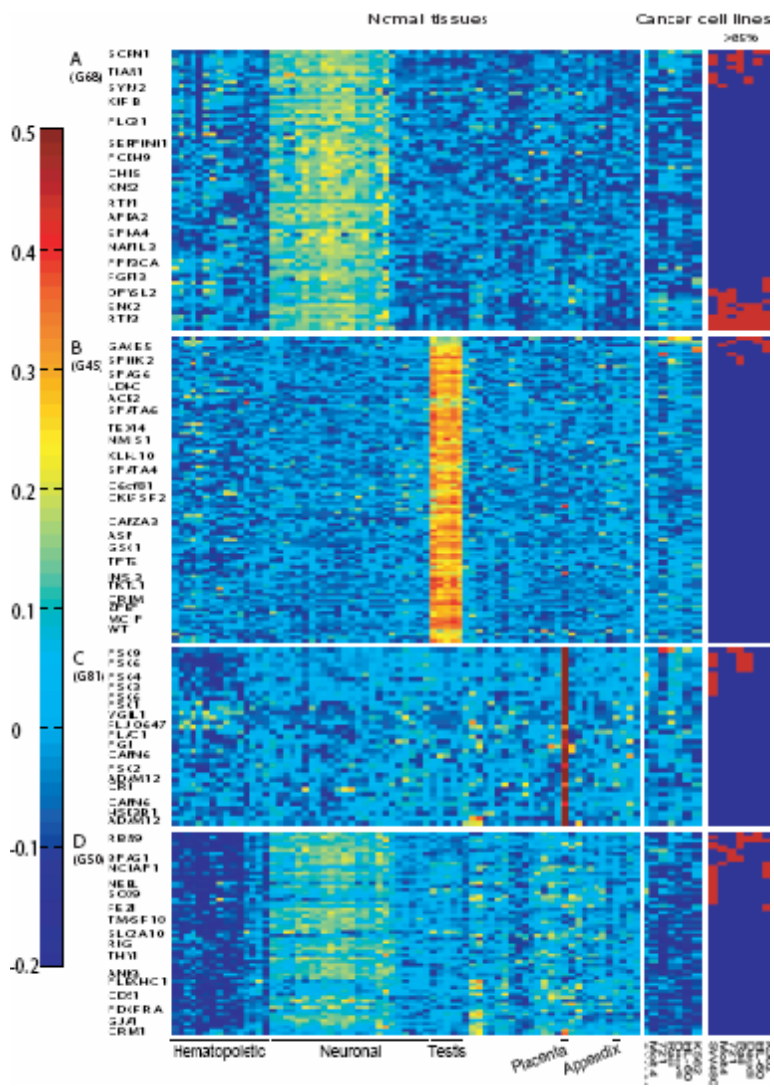

Fig 2

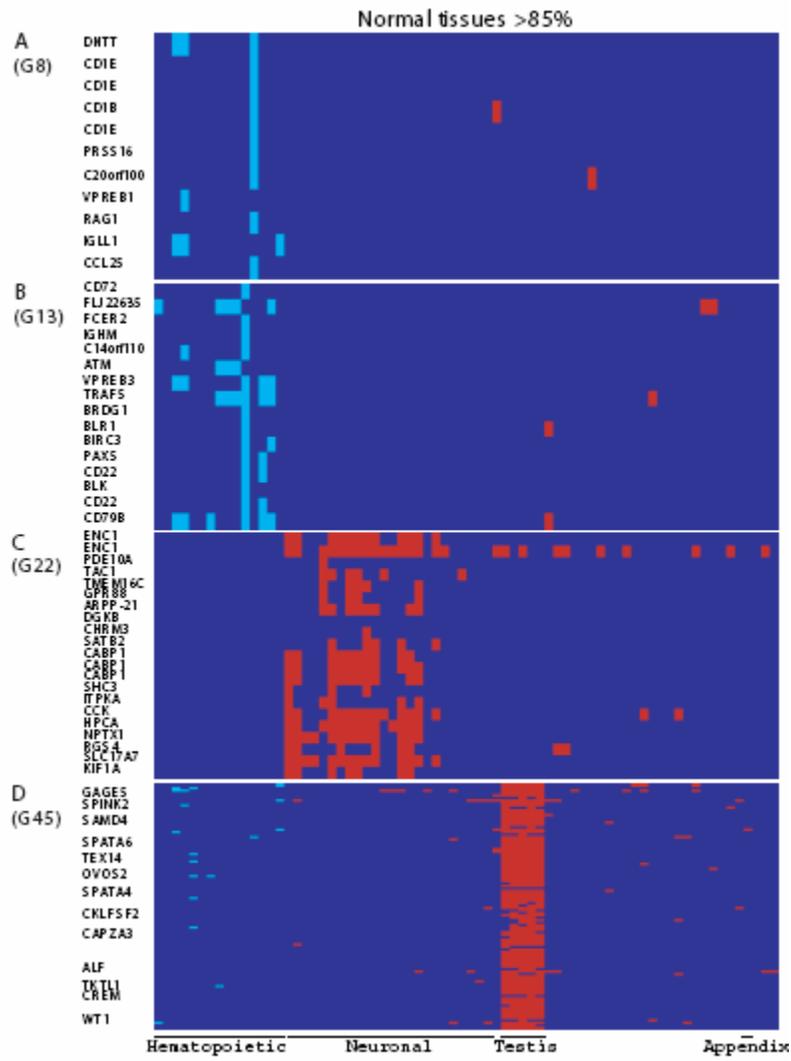

Fig 3